\documentstyle[epsf,epsfig]{l-aa}

\begin{document}

\thesaurus{10         
              (10.19.2;  
               10.19.3;  
               13.09.6)} 

\title{The bulge luminosity functions in the MSX infrared bands}


\author{M. L\'opez-Corredoira$^1$, M. Cohen$^2$, P. L. Hammersley$^1$}

\offprints{martinlc@ll.iac.es}

\institute{$^1$ Instituto de Astrof\'\i sica de Canarias,
              E-38200 La Laguna, Tenerife, Spain\\
	   $^2$ Radio Astronomy Laboratory, 601 Campbell Hall,
	   University of California, Berkeley, CA 94720, USA}

\date{Received xxxx; accepted xxxx}

\maketitle 

\markboth{MSX -- bulge}{M. L\'opez--Corredoira et al.}

\begin{abstract}
We use an inversion technique to derive the luminosity functions of the
Galactic bulge from point source counts extracted from the Midcourse
Space Experiment's Point Source Catalog (version 1.2).
\end{abstract}


\section{Introduction}
 
Information in the mid-infrared about the stars of
the bulge has, until now, been restricted to the exploitation of IRAS
data (e.g., Wainscoat et al. 1992). The new Point Source Catalogs
(Egan et al. 1999) from the 
Midcourse Space Experiment (MSX; Mill et al. 1994,
Johns Hopkins APL Technical Digest 1996, Price et
al. 1998; Price et al. 2000) provide far better spatial resolution, which 
considerably lessens the confusion due to high source density, and provides 
space-based satellite observations in six well-characterised filters.  In the 
present paper we use the MSX data to analyse the stellar populations
of the Galactic bulge in terms of their luminosity functions (LFs).

The MSX Point Source Catalogs (version 1.2, hereafter PSC1.2, obtained from
NASA/IPAC) 
contain almost one third of a million sources confined solely to the domain $|b|<5^\circ$,
providing valuable probes of the bulge. Egan et al. (1999) provide detailed
documentation for the PSC1.2.  We have, in all cases, dealt only with sources
of the highest quality (quality flags 3 or 4). These yield the most
accurate radiometry and are unconfused.

Normally, the LF is measured in small regions of sky, such as Baade's Window,
assuming equal distance for all stars, but this is a poor approximation
since it does not take into account the range in distances. 
Instead, we apply a technique developed by L\'opez-Corredoira et al. (2000) 
to deconvolve the density distribution and the luminosity function over 
wide areas that cover the bulge.  We can derive stellar statistics from a 
large sample of stars and can calculate a rather accurate LF.

\section{Luminosity functions}
Using the inversion algorithms for the bulge's stellar statistics equation
on an infrared sky survey of fields towards the centre of the Galaxy, 
it is possible to obtain information about both the LF
in the corresponding filter as well as the stellar density
distribution.  Once star counts are measured for each region, these
can be inverted, producing as a result the density distribution
along the line-of-sight and the LF.  We refer the reader interested in 
the methodology to the complete explanation given by L\'opez--Corredoira 
et al. (2000) and outlined by L\'opez-Corredoira et al. (1997, 1998), who
applied the procedure to $K$ counts drawn from the Two Micron Galactic
Survey.

\begin{figure}
\begin{center}
\vspace{1cm}
\mbox{\epsfig{file=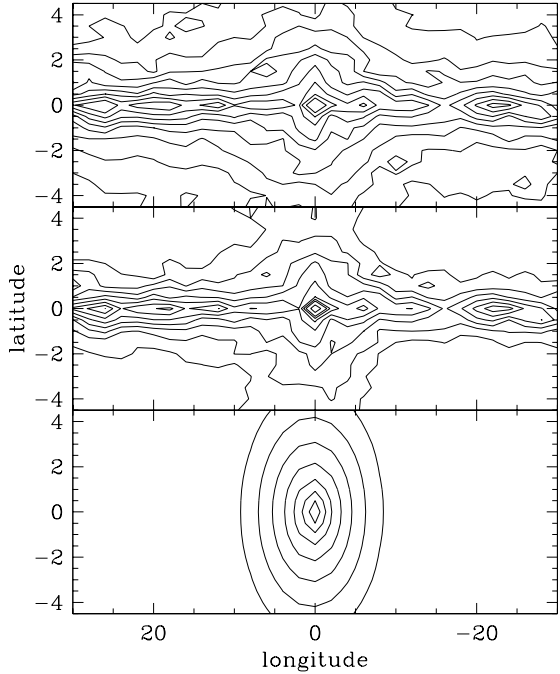,height=11cm}}
\end{center}
\caption{Up: Star counts in ``Band-A'' of MSX up to magnitude 5.4; 
contour step is 30 star/deg$^2$, first contour at 30.
Mid: Same star counts once the 
disc was subtracted; same contour step. Down: 
bulge, according to the inversion of the star
counts, projected in the sky; same contour step. As can be observed, 
the asymmetry and other features agree with the result of the inversion
in off-plane regions.}
\label{Fig:fort45}
\end{figure}

In the present paper, instead of subtracting the star counts due to the disc
using a model of the Galaxy, we evaluate it in the central regions
from an extrapolation of the star counts in outer regions.  Specifically,
we used the region $(30^\circ-|b|)<|l|<90^\circ$, 
$2^\circ<|b|<5^\circ$, where the disc is the only component 
along the line-of-sight, to evaluate the disc star counts by fitting 
bi-cubic polynomial functions in the variables of Galactic longitude
and latitude. The least-squares fits to the star
counts for each magnitude and in each filter were carried out using
the ``E02DDF'' routine of the NAG (1997) library.  
We extrapolated these functions into the central regions and 
subtracted them from the total star
counts.  By doing this, we remove the disc component and, because the
bulge is defined as the excess over the extrapolation of the disc in the
central regions of a galaxy, the star counts due to the bulge are the
natural outcome of this operation. 

We use only off-plane regions ($|b|>2^\circ$)
to avoid contamination by the spiral arms and other potential
coplanar components such as a bar or the molecular ring.
With the bulge star counts in the region within
$|l|<(12^\circ-|b|)$, $2^\circ<|b|<5^\circ$ (102 deg$^2$),
we derive the LFs for the four long-wavelength MSX bands with
isophotal wavelengths ($\lambda_{\rm iso}$) of 8.28 (``A"),
12.13 (``C''), 14.65 (``D''), and 21.33 $\mu$m (``E''). Extinction is 
neglected in this region for these bands.
The procedure could not be carried out for MSX's short-wavelength
narrow bands near 4.3 $\mu$m (``B$_1$'' and ``B$_2$'') because too few
sources are detected to make adequate fits to the disc to remove this
contaminant from the bulge by extrapolation.

The ranges of magnitude used for each of the four bands are shown in Table
\ref{Tab:filters}, along with the definition of in-band flux for zero
magnitude. The limiting magnitude for completeness in the star counts is 
the upper limit of the selected range.

\begin{table}
\begin{center}
\caption{Range of magnitudes used in the inversion for each filter
and total number of stars due to the bulge, i.e. once the disc was
subtracted, in the selected region: 
$|l|<(12\ {\rm deg.}-|b|)$, $2\ {\rm deg.}<|b|<5\ {\rm deg}$.}
\begin{tabular}{c|c|c|c|c}
Band& $\lambda_{\rm iso}$ & 0$^m$ irradiance& Range used & Bulge stars \\ 
 &  ($\mu$m)& (W cm$^{-2}$)& (mag.)& \\
\hline
A &      8.28&   8.196E-16& 2.0 to 5.4     & 4116        \\
C &     12.13&   9.259E-17& 0.0 to 3.0     & 1286        \\
D &     14.65&   5.686E-17& 0.0 to 3.0      & 1519        \\
E &     21.34&   3.555E-17&  -1.0 to 1.2     & 888 
\label{Tab:filters}
\end{tabular}
\end{center}
\end{table}

The stellar density distribution is obtained by inversion only for the
band which provides the highest number of stars, i.e. that at 8.28 $\mu$m,
and this density is used to obtain the LF in all bands by inversion.
The density distribution by inversion is a posteriori fitted to
a triaxial ellipsoid of axial ratios $1:0.36:0.23$ with centre 
8000 pc from the Sun, minor axis perpendicular to the plane, major axis
in the plane, pointing nearly to the Sun with a small inclination 
of $\sim 4$ deg. towards the first quadrant, with a density of stars

\begin{equation}
D(t)=K(t/3800)^{-0.9}\exp(-
(t/3800)^{5.7})\ {\rm star\ pc}^{-3}
\label{dens}
,\end{equation}\[
{\rm for \ }t \ {\rm between\ 1500\ and\ 4000\ pc}
\]
\[
t=\sqrt{x^2+K_y^2y^2+K_z^2z^2}
\]
where $x$, $y$, $z$ are the three axis; $K_y=2.8$, $K_z=4.4$ the two 
axial ratios (\rm{a/b} and {\rm a/c}); and $K$ is an unknown constant.
Note that the fitting distribution given in eq. (\ref{dens}) is inferred
entirely from MSX data with the procedure explained in L\'opez-Corredoira
et al. (2000) which allows to obtain both the density and the luminosity
function simultaneously.

This small-inclination model is the
best fit, sufficient to yield the observed asymmetry of the bulge, in which
the star counts at positive Galactic longitudes are higher than at negative
Galactic longitudes (see Fig. \ref{Fig:fort45}). However, because
the star count map we used is of quite low resolution
(the density of bulge stars is only
around 300 deg$^{-2}$ up to the limiting magnitude in 
the innermost bulge for Band-A; we used bins of
$(\Delta l=2^\circ)\times (\Delta b=0.5^\circ)$), and the angle
is strongly dependent on small variations in the position of
the maximum projected star counts, we regard these numbers as 
a rough approximation.

The structural information is not totally reliable because we
are using only a few thousand stars, insufficient for a statistically 
adequate inversion of the density distribution. However, as argued 
by L\'opez-Corredoira et al. (1997, 2000), the inversion to
obtain the luminosity function is much less sensitive to Poissonian noise, 
and only weakly dependent on the density distribution because the dispersion in
distance for bulge stars is relatively small. Therefore, our conclusions
about the luminosity function are much more robust.

\begin{figure*}
\begin{center}
\vspace{1cm}
\mbox{\epsfig{file=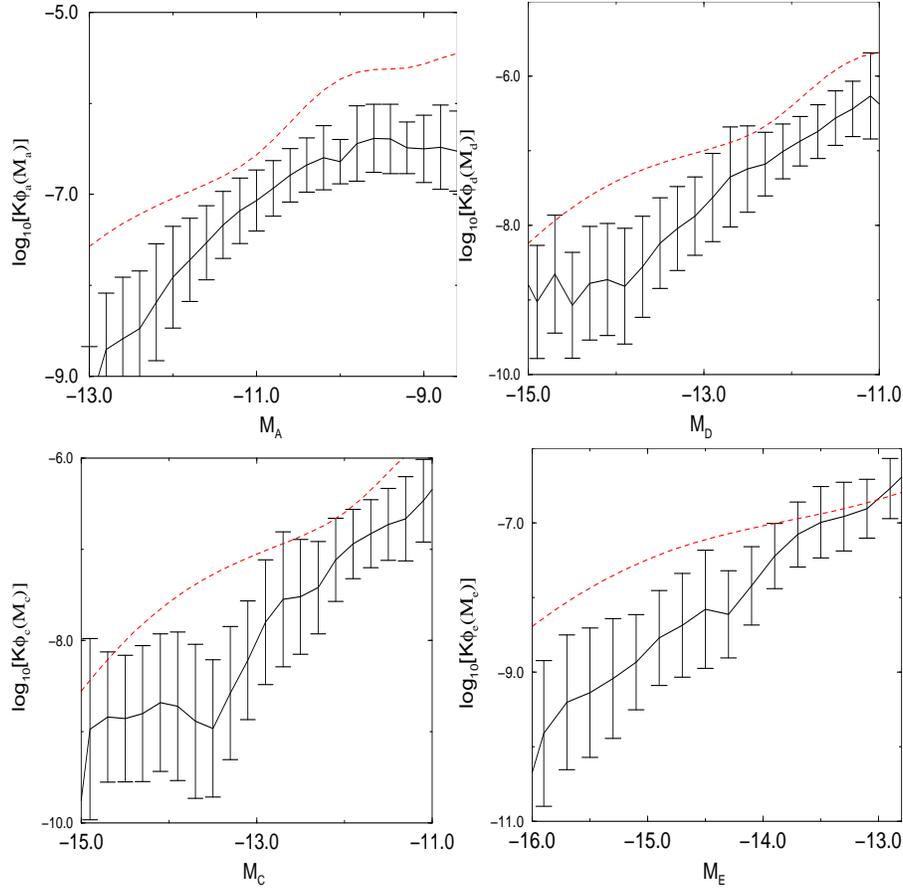,height=12cm}}
\end{center}
\caption{Luminosity functions for bulge stars in the four long-wavelength MSX bands
(solid line): A, C, D, E. $\phi $ is the normalized luminosity function, and 
$K$ is an unknown constant. Dashed lines show for comparison
the normalized luminosity
functions of ``SKY'' model ($K=1$).}
\label{Fig:PHI}
\end{figure*}

The LFs of the bulge stars for the four filters are
shown in Fig. \ref{Fig:PHI}. The luminosity function for other
bands than A are derived using 
the inversion procedure by L\'opez-Corredoira et al. (2000), but in these
cases using a given density (\ref{dens}) to obtain the luminosity function
instead of obtaining both the density and the luminosity function.
The range of absolute magnitudes depends
on the selected range in apparent magnitudes for each filter, 
and on the range of distances to the bulge stars. The error bars 
come from the dispersion of LFs in the different directions.

\section{Comparison with ``SKY'' model}

For comparison, we have also plotted in Figure \ref{Fig:PHI}
the normalized 
LF of the ``SKY'' model (Cohen 1994, 1995;
Cohen, Sasseen \& Bowyer 1994), which is a modified version of
the model introduced by Wainscoat et al. (1992), but one whose changes do not
significantly affect the bulge. To assemble the requisite data within SKY
we first ran the model in its customised mode and generated tables of absolute
magnitudes for all 87 categories of celestial source. These were then
combined with the bulge mask (cf. Wainscoat et al. 1992) and the newest
space densities.

Note that figure \ref{Fig:PHI} compares SKY and MSX LFs only to show the differences 
in slope, rather than in absolute values; i.e., in each plot, the SKY LF shown    
corresponds to $K=1$ but the LF derived from MSX data has been
arbitrarily normalized ($K\ne 1$) for the comparisons.  The differences in
slopes of the MSX plots compared to those of the SKY model are noteworthy, 
especially for Band-E, although one should recognize the uncertainties
resulting from these inversions.
The longer the filter wavelength, the steeper the slope of the MSX LF in 
comparison with those of the SKY model, as shown in Table \ref{Tab:slopes},
where the slope is $s_\lambda $ such that

\begin{equation}
\phi _\lambda (M_\lambda )\approx C_\lambda \times 10^{s_\lambda M_\lambda}
\end{equation}
in the range where we have calculated the LFs.
The constants, $C_\lambda $, depend on the normalizations.
The slopes are dependent on the magnitude range, but the ratio shown
in the fourth column of the table should not be.
 
\begin{table*}
\begin{center}
\caption{Average slopes of the plots in Figure \protect{\ref{Fig:PHI}}.}
\begin{tabular}{c|c|c|c}
Band  & MSX Slope & SKY slope & Ratio of slopes (MSX/SKY) \\ \hline
A & $0.56\pm 0.05$ & $0.52\pm 0.02$ & $1.08\pm 0.11$ \\
C & $0.74\pm 0.05$ & $0.61\pm 0.02$ & $1.21\pm 0.09$ \\
D & $0.78\pm 0.03$ & $0.60\pm 0.02$ & $1.30\pm 0.07$ \\
E & $1.08\pm 0.03$ & $0.51\pm 0.03$ & $2.12\pm 0.14$
\label{Tab:slopes}
\end{tabular}
\end{center}
\end{table*}

We can also compare the absolute counts. If we integrate the LFs
over the whole volume of the bulge (see \S 6.5 of 
L\'opez-Corredoira et al. 2000), and multiply it by the total density of 
stars given by (\ref{dens}), assuming it is valid for all $t$, 
we obtain the results shown in table
\ref{Tab:abscounts}, by comparison with the SKY model predictions.
There is another clear trend, namely the increase, with increasing wavelength,
in the ratio of the observed numbers of stars to those predicted. 
The absolute values of the ratios themselves are not important, 
since we used the approximation that (\ref{dens}) is valid for
all $t$, but the trend of increasing ratio highlights a significant 
difference in stellar population with respect to that of the SKY model,
which will be valuable for future refinements of SKY.

\begin{table*}
\begin{center}
\caption{Number of stars ($n$) in the whole bulge for a limiting range
of magnitudes.}
\begin{tabular}{c|c|c|c|c}
Band  & Range of magnitudes & $n$ from MSX & $n$ from SKY 
& Ratio of $n$ (MSX/SKY) \\ \hline
A & $M_A<-8.6$ & $3.4\times 10^4$ & $7.8\times 10^4$ & 0.44\\
C & $M_C<-11.0$ & $1.4\times 10^4$ & $1.9\times 10^4$ & 0.74 \\
D & $M_D<-11.0$ & $1.8\times 10^4$ & $2.7\times 10^4$ & 0.67 \\
E & $M_E<-12.8$ & $1.0\times 10^4$ & $0.46\times 10^4$ & 2.17
\label{Tab:abscounts}
\end{tabular}
\end{center}
\end{table*}

Bulge stars that contribute to the counts in our adopted ranges of 
magnitude are basically AGBs for Band-A (with absolute
magnitudes around -9). 
The reddest AGBs are those obscured by the thickest dust
shells while those nearer the faint limit of our adopted magnitude range have 
lower mass-loss rates. Since the slope of the luminosity
function and the absolute number of stars are larger in MSX than predicted by
SKY for larger wavelengths, we are led to the following suggestions about SKY:
there is a deficit of weak mass loss AGB stars in ``SKY", for 
example, the objects discussed by Glass et al. (1999).

Source densities in the bulge are certainly high and, at densities above
500 sources deg$^{-2}$ in filter A, confusion is sufficient to limit the
completeness of PSC1.2 (Egan et al. 1999; their \S7.1).
PSC1.2 contains about 3300 confused sources over 
all longitudes for latitudes in the plane with modulus exceeding 2$^\circ$
(Egan et al. 1999; their Table 10).  Therefore, we expect a small
underestimation of the total number of stars in our two selected 
bulge latitude ranges.  Incompleteness in the differential counts from
PSC1.2 can be seen in the inner Galaxy at around 200 mJy in filter A
(Egan et al. 1999; their Figs. 31, 38), equivalent to $M_A\sim-8.5$.  
Thus, the combination of confusion and incompleteness might furnish an
adequate explanation for the ``flattening" of our derived LF at faint 
8.28-$\mu$m magnitudes (Fig. \ref{Fig:PHI}) and the small ratio of {\rm n}(MSX)/{\rm n}SKY
in Table 3 although we are not sure about this to be the reason
since confusion effects should be small.

MSX's band A is more than an order of magnitude more sensitive than the
other bands. Consequently, we would expect the depth of coverage of the
bulge in bands CDE to be comparable to that of IRAS, and hence we expect
few differences in completeness between MSX CDE and IRAS at 12 or 25 $\mu$m.
Hence, we dismiss the scenario of an excess of AGBs with thick 
dust envelopes (strong mass loss).

The SKY model was originally guided by IRAS mid-infrared data although
it has always been capable of handling the prediction of counts from $BVJHK$  
filters, as well as IRAS's, and from user-supplied arbitrary passbands
between 2.0 and 35.0 $\mu$m, hence its capability to operate equally well
for these MSX filters.  As Wainscoat et al. (1992) demonstrated, SKY
fits the IRAS source counts well. The very red AGBs detected by MSX are 
also detected by IRAS, although not many of those with weak mass loss. 
It is possible that the pattern of excesses and deficits in LFs that we have
described are compensated for, in predicting IRAS counts, but not for MSX 
whose deeper counts can probe specific populations better. The improvements
delineated here will be applied to a future version of the SKY model.

The LFs established in this paper may be used to
fit the population of bulge stars in any model. We do not pursue
these fits here because it is not the aim of the present paper.

\section{Conclusions}

The LFs of the stellar population of the Galactic bulge
in four infrared filters --- 8.28, 12.13, 14.65 and 21.33 $\mu $m --- 
have been calculated by means of a special technique of inversion.
Based on the new MSX data, our inversion provides this information
for the first time in the literature. It is of general value
in the study of these stellar populations and will be implemented
in a future version of the SKY model.

\begin{acknowledgements}

Thanks are given to the anonymous referee for helpful comments which have
led to significant improvements in the paper.
The MSX Point Source Catalogs were obtained
from the NASA/IPAC Infrared Science Archive at Pasadena, California.

\end{acknowledgements}

\end{document}